\documentclass{iopart}
\usepackage{iopams,amsfonts,amssymb,amsthm} 
\usepackage{geometry}                
\usepackage{etex}
\usepackage[pdftex]{graphicx}
\usepackage{rawfonts}
\usepackage{amssymb}
\usepackage{epstopdf}
\usepackage[usenames,dvipsnames,svgnames]{xcolor}
\expandafter\let\csname equation*\endcsname\relax
\expandafter\let\csname endequation*\endcsname\relax
\usepackage{amsmath}
\newcommand{\avm}{\langle v \rangle}
\newcommand{\avh}{\langle h \rangle}

\newcommand{\scw}{\mathcal{W}}
\newcommand{\fsfree}{\omega_t^{(n)}}

\input prepictex.tex
\input pictex.tex
\input postpictex.tex

\newtheorem{lemm}{Lemma}

\def\Pr{\noindent \emph{Proof: }}
\def\qed{$\Box$}

\def\nor{\normalsize}

\def\sfrac#1#2{\hbox{\nor $\frac{#1}{#2}$}}
\def\Sfrac#1#2{\hbox{\large $\frac{#1}{#2}$}}

\def\shalf{{\sfrac{1}{2}}}

\def\Ref#1{(\ref{#1})}

\usepackage{hyperref}
\hypersetup{colorlinks,citecolor=blue,filecolor=blue,linkcolor=blue,urlcolor=blue}

\begin{document}

\title{Adsorbed self-avoiding walks pulled at an interior vertex}
\author{C J Bradly$^*$, E J Janse van Rensburg$^{\dagger}$, A L Owczarek$^*$ and S G  Whittington$^\ddagger$ }
\address{
{}$^*$School of Mathematics \& Statistics, University of Melbourne, Victoria 3010, Australia \\
{}$^\dagger$Department of Mathematics \& Statistics, York University, Toronto M3J 1P3, Canada \\
{}$^\ddagger$Department of Chemistry, University of Toronto, Toronto M5S 3H6, Canada 
}

\begin{abstract}
We consider self-avoiding walks terminally attached to a surface at which they 
can adsorb.  A force is applied, normal to the surface, to desorb the walk and we 
investigate how the behaviour depends on the vertex of the walk at which the 
force is applied.  We use rigorous arguments to map out some features of the phase diagram, 
including bounds on the locations of some phase boundaries, and we use Monte
Carlo methods to make quantitative predictions about the locations of these 
boundaries and the nature of the various phase
transitions.
\end{abstract}

\maketitle

\section{Introduction and review}
\label{sec:Introduction}

Polymer adsorption at a surface has been studied for many years \cite{Rubin1965}.  
More recently, with 
the introduction of techniques such as atomic force microscopy (AFM), it is possible to pull
an adsorbed polymer off a surface and measure the required critical force for desorption
\cite{Haupt1999,Zhang2003}.  This has led to a renewed interest in how polymers respond to 
an applied force \cite{Beaton2015,GuttmannLawler,IoffeVelenik,IoffeVelenik2010,Rensburg2016,Orlandini2016}.

Self-avoiding walk (SAW) models of polymers \cite{Rensburg2015,MadrasSlade} 
adsorbed at a surface and desorbed by the action of a force have been investigated 
previously.  Most of the available results are about the case where the walk is terminally 
attached to an impenetrable surface and where the force is applied at the other unit 
degree vertex \cite{Guttmann2014,Rensburg2013,
Krawczyk2005,Krawczyk2004,Mishra2005}.  For a directed version of this model,
see reference \cite{Rensburg2011} and for related work see
\cite{Skvortsov2009,Binder2012}.

For the $d$-dimensional hypercubic lattice $\mathbb{Z}^d$ let  the vertices 
have coordinates $(x_1,x_2, \ldots x_d)$, $x_i \in \mathbb{Z}$.
If $c_n$ is the number of $n$-edge self-avoiding walks starting at the origin
then \cite{Hammersley1957}
\begin{equation}
\log d \le \lim_{n\to\infty} \sfrac{1}{n} \log c_n = \inf_{n>0} \sfrac{1}{n} \log c_n = \log \mu_d \le \log(2d-1)
\end{equation}
where $\mu_d$ is the \emph{growth constant} of the self-avoiding walk. 
If the walk is constrained to lie in the half-lattice with $x_d\geq 0$ while its
first vertex is attached to the origin in the hyperplane $x_d=0$ (the
\textit{adsorbing plane}), then it is a \emph{positive walk} and we write 
$c_n^+$ for the number of $n$-edge positive walks.    It is known 
\cite{Whittington1975} that $\lim_{n\to\infty} \sfrac{1}{n} \log c_n^+ = \log \mu_d$.

Let $c_n^+(v,h)$ be the number
of $n$-edge positive walks with $v+1$ vertices in the hyperplane 
$x_d=0$ and with the $x_d$-coordinate of the last vertex equal to 
$h$.  We say that the walk has $v$ \emph{visits} and the last vertex
has \emph{height} equal to $h$.  Define the partition function
\begin{equation}
C_n^+(a,y) = \sum_{v,h} c_n^+(v,h) a^v y^h,
\end{equation}
where $a=\exp(-\epsilon/k_\text{B}T)$ and $y=\exp(F/k_\text{B}T)$ are the Boltzmann weights associated with the monomer-surface interaction energy $\epsilon$ and the pulling force $F$, respectively.  If $F>0$ then $y>1$ and the force is directed away from the surface.

Suppose $y=1$ so that the positive walk interacts with the surface but is not subject to a force.
This is the pure adsorption problem.  The free energy is 
\begin{equation}
\kappa(a) = \lim_{n\to\infty} \sfrac{1}{n} \log C_n^+(a,1)
\end{equation}
and $\kappa(a)$ is a convex function of $\log a$ \cite{HTW}.
There exists a critical value of $a$, $a_c > 1$,
such that $\kappa(a) = \log \mu_d$ when $a \le a_c$ and 
$\kappa(a) > \log \mu_d$ when $a > a_c$, so
that $\kappa(a)$ is singular at $a=a_c > 1$ \cite{HTW,Rensburg1998,Madras2017}.

If  $a=1$ the walk does not interact with the adsorbing plane and the
free energy is
\begin{equation}
\lambda(y) = \lim_{n\to\infty} \sfrac{1}{n} \log C_n^+(1,y).
\end{equation}
$\lambda(y)$ is singular at $y=1$ \cite{Beaton2015,IoffeVelenik,IoffeVelenik2010} and
the walk is in a ballistic phase when $y > 1$.   It is also a convex function of 
$\log y$ \cite{Rensburg2009}.

In the general situation where $a \ne 1$ and $y \ne 1$ there is a thermodynamic
limit in the model and the free energy is given by \cite{Rensburg2013} 
\begin{equation}
\psi(a,y) = \lim_{n\to\infty} \sfrac{1}{n} \log C_n^+(a,y)
= \max[\kappa(a), \lambda(y)].
\label{eq:psicondition}
\end{equation}
In particular, $\psi(a,y) = \log \mu_d$ when $a \le a_c$
and $y \le 1$.  For $a > a_c$ and $y > 1$ there is a phase
boundary in the $(a,y)$-plane along the curve 
given by $\kappa(a) = \lambda(y)$.  This phase transition
is first order \cite{Guttmann2014}.  The fact that the phase boundary is determined
by the condition that $\kappa(a) = \lambda(y)$ has been used to locate the 
phase boundary accurately using exact enumeration and series analysis
\cite{Guttmann2014}. 









If atomic force microscopy is used to pull the adsorbed polymer off the surface it is possible
to apply the force at the last monomer (by functionalizing that monomer and attaching it to the
AFM tip by a covalent bond).  More typically the tip is brought into contact with the polymer and the 
force might be applied at any monomer \cite{Zhang2003}.  
This raises the question of how the critical force
for desorption depends on where the force is applied \cite{Rensburg2017}.

\begin{figure}[t!]
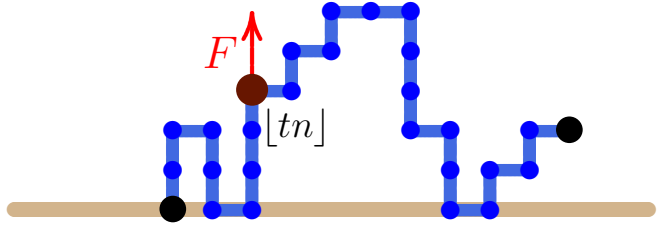

\beginpicture
\setcoordinatesystem units <1.5pt,1.5pt>
\setplotarea x from -30 to 100, y from  0 to 55
\setplotarea x from 0 to 100, y from  0 to 55

\setcoordinatesystem units <1.5pt,1.5pt> point at -40 0 
\color{Tan}
\setplotsymbol ({\Large$\bullet$})
\plot -30 0 130 0  /

\color{red}
\setplotsymbol ({\LARGE.})
\arrow <10pt> [.2,.67] from 30 30  to 30 50 
\put {\LARGE$F$} at 22 40

\color{black}
\put {\Large$\lfloor tn\rfloor$} at 41 21 

\setplotsymbol ({\large$\bullet$})
\color{RoyalBlue}

\plot 10 0 10 10 10 20 20 20 20 10 20 0 30 0 30 10 30 20 30 30 40 30 40 40   /
\plot 40 40 50 40 50 50 60 50 70 50 70 40 70 30 70 20 80 20 80 10 80 0 90 0 /
\plot 90 0 90 10 100 10 100 20 110 20 /

\color{black}
\multiput {\scalebox{1.25}{{\huge$\bullet$}}} at 10 0 110 20  /
\color{blue}
\multiput {\scalebox{1.25}{{\Large$\bullet$}}} at  10 10 10 20 20 20 20 10 20 0 30 0 
30 10 30 20 30 30 40 30 40 40 50 40 50 50 60 50 70 50 70 40 70 30 70 20 80 20 80 10 
80 0 90 0 90 10 100 10 100 20 /
\color{Sepia}
\multiput {\scalebox{1.5}{{\huge$\bullet$}}} at 30 30  /

\color{black}
\normalcolor

\endpicture
\caption{An adsorbing self-avoiding walk of length $n$ attached to
the adsorbing plane and being pulled by a force $F$ at a vertex which is
a chemical distance $\lfloor tn \rfloor$ from the origin.  This walk has $4$
visits and the height of the pulled vertex is $3$.}
\label{fig:forcesketch}
\end{figure}

In this paper we consider the case where the force is applied at a vertex
which is a chemical distance $\lfloor tn \rfloor$ from the origin.  This is a model
of an attached adsorbing linear polymer being pulled at a vertex which is a distance
$tn$ along the polymer by a vertical force $F$ (see Fig. \ref{fig:forcesketch}).

Number the vertices along the self-avoiding walk $j=0,1,\ldots n$
where the zero'th vertex is at the origin.  If the force is applied at 
the vertex numbered $\lfloor tn \rfloor$ and $t \ge \sfrac{1}{2}$, then it is 
known \cite{Rensburg2017}  that the phase diagram is similar to the case where $t=1$ and
the force is applied at the unit degree end-vertex of the walk.
Less is known if $0<t<\sfrac{1}{2}$, but it is established that there is an
additional \emph{mixed phase} where the free energy depends on both $a$ and $y$ 
and that the free energy is a function of $t$, namely the point where the force
is applied \cite{Rensburg2017}. 

In section \ref{sec:rigorous} we examine the phase diagram of the model when $0 < t < 1/2$.
We prove that there are four phases in the model, namely a \emph{free phase},
an \emph{adsorbed phase}, a \emph{ballistic phase}, and a \emph{mixed phase},
and we obtain rigorous bounds on the locations of the boundaries between the 
ballistic and mixed phases and between the mixed and adsorbed phases.
In section \ref{sec:montecarlo} we employ Monte Carlo simulation using the flatPERM algorithm to investigate these results for finite-size walks.
We simulate SAWs of length $n=256$ for several values of $t\leq 1/2$ on the two-dimensional square lattice and three-dimensional simple cubic lattice which are expected to produce qualitatively similar results.
For the two-dimensional case we also use exact enumeration data from Ref.~\cite{Guttmann2014} to visualise results for the phase boundaries in comparison to the Monte Carlo results.

\section{Rigorous results}
\label{sec:rigorous}

Consider an $n$-edge self-avoiding walk attached at its zero'th vertex to an 
impenetrable plane $x_d=0$, with $v+1$ vertices in this plane and having 
the $x_d$-coordinate of its vertex numbered $\lfloor tn \rfloor$ equal to $h$.  We say
that the vertex has \emph{height} $h$ and that the walk has $v$ \emph{visits}.  
We shall be concerned with the case where the force is applied at 
vertex numbered $\lfloor tn \rfloor$ and $0 < t \le 1/2$ (see Fig. \ref{fig:forcesketch}).

Write $w_n^{(t)}(v,h)$ for the number of such walks and write the partition function as
\begin{equation}
W_n^{(t)}(a,y) = \sum_{v,h} w_n^{(t)}(v,h)a^vy^h.
\label{eq:PartitionFunction}
\end{equation}
When we can prove that the limit exists we shall write $\omega_t(a,y) 
=\lim_{n\to\infty} \sfrac{1}{n} \log W_n^{(t)}(a,y)$ for the free energy.

First we prove a result about the free energy when $0 < t < 1$, 
$y > 1$ and $a \le a_c$.  This extends a result in \cite{Rensburg2017}
when $y>1$ and $a \le 1$.
\begin{lemm}
When $0 < t < 1$, $y>1$ and $a\le a_c$ the free energy is given by
$$\omega_t(a,y) = t \lambda(y) + (1-t) \log \mu_d.$$
\label{lemm:ballistic}
\end{lemm}
\Pr
We know from \cite{Rensburg2017} that $\omega_t(a,y) = t \lambda(y) + (1-t) \log \mu_d =\omega_t(1,y)$
when $a \le 1$ and $y > 1$.  For $a \le a_c$ and $y > 1$ monotonicity implies that the 
free energy is bounded below by $\omega_t(1,y)$.  To obtain an upper bound recall
that the walk is subject to a force at vertex $\lfloor tn \rfloor$.  Either the walk returns to 
the adsorbing plane after vertex $\lfloor tn \rfloor$ or it does not return after vertex $\lfloor tn \rfloor$.  
In the latter case subdivide the walk into two subwalks at
vertex $\lfloor tn \rfloor$.  The first subwalk has extensive free energy 
$\lfloor tn \rfloor \lambda(y) + o(n)$ and the second subwalk has extensive
free energy equal to $(n-\lfloor tn \rfloor) \log \mu_d + o(n)$.  Treating the two walks as 
independent, adding the two terms together, dividing by $n$ and letting $n \to \infty$,
 shows that $t \lambda(y) + (1-t) \log \mu_d$ is an upper bound on the
(intensive) free energy.  If the walk does return to the adsorbing surface 
after vertex $\lfloor tn \rfloor$, suppose that the first 
return is at vertex $\lfloor sn \rfloor$.  Subdivide the walk into three subwalks at
vertex $\lfloor sn \rfloor -1$ and at vertex $\lfloor sn \rfloor$.  For the first 
walk, with $\lfloor sn \rfloor - 1$ edges, 
we have a condition that the last vertex is at a distance 1 above the surface.  
The free energy is bounded
above by that of the set of walks where this last vertex is at any positive distance above the surface.
The free energy of the first subwalk is therefore bounded above by
$$t \max[\kappa(a), \lambda(y)]  + (s-t) \log \mu_d = t \lambda(y) + (s-t) \log \mu_d$$
since $a \le a_c$ and $y > 1$ so $\lambda(y) > \kappa(a) = \log \mu_d$.  The middle subwalk 
has exactly one edge and one vertex in the surface and so makes no contribution to the 
free energy (after dividing by $n$ and letting $n \to \infty$).   The final subwalk has free 
energy $(1-s) \kappa(a) = (1-s) \log \mu_d$ since $a \le a_c$.  Treating the three walks as
independent and adding their contributions gives the upper bound 
$$ t \lambda(y) + (s-t) \log \mu_d + (1-s) \log \mu_d = t \lambda(y) + (1-t) \log \mu_d$$
which completes the proof.
\qed

Lemma \ref{lemm:ballistic} shows that
the walk is in a \emph{ballistic phase} when $a \le a_c$ and $y > 1$.  The free energy is then
$\omega_t(a,y) = t \lambda(y) + (1-t) \log \mu_d$.

Next we shall state some results that were proved in \cite{Rensburg2017}.  
\begin{enumerate}
\item
When $a \le a_c$ and $y \le 1$, $\omega_t(a,y) = \log \mu_d$.  We say that the 
walk is in a \emph{free phase}.
\item
When $y \le 1$, $\omega_t(a,y) = \kappa(a)$.  If $y \le 1$ and $a >a_c$ the walk
is in an \emph{adsorbed phase}.
\end{enumerate}

In addition it was proved in \cite{Rensburg2017} that, when $0 < t < 1/2$,
\begin{equation}
\liminf_{n\to\infty} \sfrac{1}{n} \log W_n^{(t)}(a,y) \ge 
\max [t \lambda(y) + (1-t) \log \mu_d,\chi(a,y),\kappa(a) ],
\label{eq:lowerbounds}
\end{equation}
where $\chi(a,y) = 2t \lambda(\sqrt{y}) + (1-2t) \kappa(a)$.
The first term corresponds to the free energy in the ballistic phase, the third to the free 
energy in the adsorbed phase and the second term  is a lower bound on the free energy
in the mixed phase.  This expression was used in \cite{Rensburg2017} to prove
that a mixed phase exists for all $0<t<1/2$.

Now we examine where pairs of these bounds become equal.  The condition
\begin{equation}
t\lambda(y) +(1-t) \log \mu_d = 2t \lambda(\sqrt{y}) + (1-2t) \kappa(a)
\label{eq:conditionI}
\end{equation}
defines a curve $y = y^I(a)$ in the $(a,y)$-plane, and the condition
\begin{equation}
2t\lambda(\sqrt{y})  + (1-2t) \kappa(a) = \kappa(a)
\end{equation}
or, equivalently,
\begin{equation}
\lambda(\sqrt{y})  = \kappa(a)
\label{eq:conditionII}
\end{equation}
defines a curve $y = y^{II}(a)$ in the $(a,y)$-plane.  Note that $y^I(a_c) = y^{II}(a_c) = 1$
so both curves pass through the point $(a_c,1)$.  

%
In the next Lemma we address the monotonicity of $y^I(a)$ and $y^{II}(a)$.  
\begin{lemm}
When $a > a_c$ the functions $y^I(a)$ and $y^{II}(a)$ are monotone increasing functions of $a$.
\label{lemm:monotone}
\end{lemm}
\Pr
Rewrite (\ref{eq:conditionI}) as
\begin{equation}
t[\lambda(y) - 2 \lambda(\sqrt{y})] = (1-2t) \kappa(a) - (1-t) \log \mu_d.
\label{eq:newcondition}
\end{equation}
The log-convexity of 
$\lambda(y)$ implies that $\lambda(y) - 2 \lambda(\sqrt{y})$ has positive derivative
a.e. for all $y >1$ so both the left hand side and right hand side of 
(\ref{eq:newcondition}) are monotone increasing functions.
This shows that $y=y^I(a)$ is monotone increasing.  A similar argument shows that 
$y^{II}(a)$ is monotone increasing.  This follows directly from the monotonicity 
of $\kappa(a)$ and $\lambda(\sqrt{y})$.
\qed

We can look at the behaviour at large $y$ by using our knowledge of the asymptotics of $\lambda(y)$.
We know that $\lambda(y) \to \log y$ for large enough values of $y$
\cite{Rensburg2013}.  In this asymptotic regime we can substitute $\lambda(y) = \log y$ in (\ref{eq:conditionI}) and solve giving
\begin{equation}
\kappa(a)=\left( \frac{1-t}{1-2t} \right)\log \mu_d.
\label{eqn13}
\end{equation}
Since $\kappa(a)$ is continuous and monotone strictly increasing for $a>a_c$, this equation
has a solution $a_0(t)$ for every $t\in (0,\sfrac{1}{2})$. 
Since $\kappa(a)$ is a strictly increasing function of $a$ for $a > a_c$, and the right
hand side of equation \Ref{eqn13} is a strictly increasing function of $t$ for
$t\in (0,\sfrac{1}{2})$, 
\begin{equation}
a_0(t)= \kappa^{-1}\left(   \frac{(1-t)\log \mu_d}{1-2t}   \right)
\label{eq:a0Definition}
\end{equation}
is a strictly increasing function of $t$ for $0<t<1/2$.   As $t \to 0$
$a_0(t) \to a_c$ because $\kappa(a)$ is strictly monotone increasing
for $a > a_c$  Similarly, as $t \to \shalf$ from below, $a_0(t)$ diverges.

In a similar way we can insert the asymptotic forms  $\lambda(y) \to \log y$  \cite{Rensburg2013}
and $\kappa(a) \to \log a + \log \mu_{d-1}$ \cite{RychlewskiJSP} in
(\ref{eq:conditionII}).  This implies that, in the asymptotic regime, 
$y^{II}(a) \sim \mu_{d-1}^2a^2$.

We write $y=y^{BM}(a)$ for the ballistic-mixed phase boundary and $y=y^{MA}(a)$
for the phase boundary between the mixed and adsorbed phases, and we next address 
the connection between these two phase boundaries and the two curves $y^I(a)$ 
and $y^{II}(a)$.  Suppose that $a > a_c$ and $y>1$.
If $y > y^{II}(a)$ the free energy is strictly greater than
$\kappa(a)$ and the system is not in the adsorbed phase.  Similarly, if
$y < y^I(a)$ the system is not in the ballistic phase.  If $y^I(a) > y^{II}(a)$ there is a region
of the $(a,y)$-plane where the system is not in either the adsorbed or ballistic phases 
and we have the inequalities $y^{MA}(a) \le y^{II}(a) < y^I(a) \le y^{BM}(a)$.  If 
$y^I(a) < y^{II}(a)$ these conditions are not met.  We make this explicit in the next
two Lemmas.

\begin{lemm}
When $a > a_0(t)$, $y^{MA}(a) \le y^{II}(a)$.
\label{lemm:boundary1}
\end{lemm}
\Pr
Since the curve $y=y^I(a)$ is asymptotic to $a=a_0(t)$ and lies to the left of this line,
$y^{II}(a)$ and $y^I(a)$ cannot intersect beyond $a = a_0(t)$.  Beyond this point we have 
$y^{MA}(a) \le y^{II}(a)$.\qed

The curve $y=y^{II}(a)$ intersects the line $a=a_0(t)$ when 
$\lambda(\sqrt{y})=\left[\sfrac{1-t}{1-2t}\right] \log \mu_d$.   
The solution of this equation is 
\begin{equation}
y=y_0(t) = \left(\lambda^{-1}\left(\frac{(1-t) \log \mu_d}{1-2t} \right)   \right)^2
\end{equation}
because $\lambda(y)$ is strictly monotone for $y > 1$.
This implies the following:
\begin{lemm}
When $y > y_0(t)$, $y^{BM}(a) \ge y^{I}(a) > y^{II}(a)$.
\label{lemm:boundary2}
\end{lemm}

These results imply that the phase boundary $y=y^{BM}$  between the 
ballistic and mixed phases lies between the line $a=a_c$ and the line $a=a_0(t)$.  
In addition, at large values of $y$,
the phase boundary $y=y^{MA}$ between the mixed and adsorbed phases cannot increase 
more rapidly than quadratically in $a$.

We can look at this from another point of view.  Suppose that $a^I(y)$ and $a^{II}(y)$
are the inverse functions to $y^I(a)$ and $y^{II}(a)$.  Then $y^I(a) > y^{II}(a)$ 
implies that $a^{II}(y) > a^I(y)$.  
But 
\begin{equation}
a^I(y) = \kappa^{-1} \left( \frac{t\lambda(y) + (1-t)\log \mu_d - 2t \lambda(\sqrt{y})}{1-2t} \right)
\end{equation}
and
\begin{equation}
a^{II}(y) = \kappa^{-1} ( \lambda(\sqrt{y})).
\end{equation}
Since $\kappa(a)$ is monotone increasing the condition $a^{II}(y) > a^I(y)$
is equivalent to 
\begin{equation}
\lambda(\sqrt{y}) > \frac{t\lambda(y) + (1-t)\log \mu_d - 2t \lambda(\sqrt{y})}{1-2t}
\end{equation}
or, equivalently,
\begin{equation}
\lambda(\sqrt{y}) > t\lambda(y) + (1-t)\log \mu_d.
\end{equation}
Since $\lambda(y) \to \log y$ this condition is always satisfied for $0 < t < 1/2$
at sufficiently large $y$.  This gives an alternative proof that there is a 
mixed phase for all $0 < t < 1/2$.

We have proved the existence of four phases, free, adsorbed, ballistic and mixed.  
However, we cannot establish rigorously the order of the ballistic-mixed or adsorbed-mixed phase transitions.
This is because we only have a lower bound on the free energy in the mixed phase.  
Although we know that a mixed phase exists for all $t < 1/2$ we do not know rigorously whether the mixed phase extends down to $(a_c,1)$ or whether there is a phase boundary between the ballistic and adsorbed phases for $a$ close to $a_c$ and $y$ close to 1.

\begin{figure}[t!]
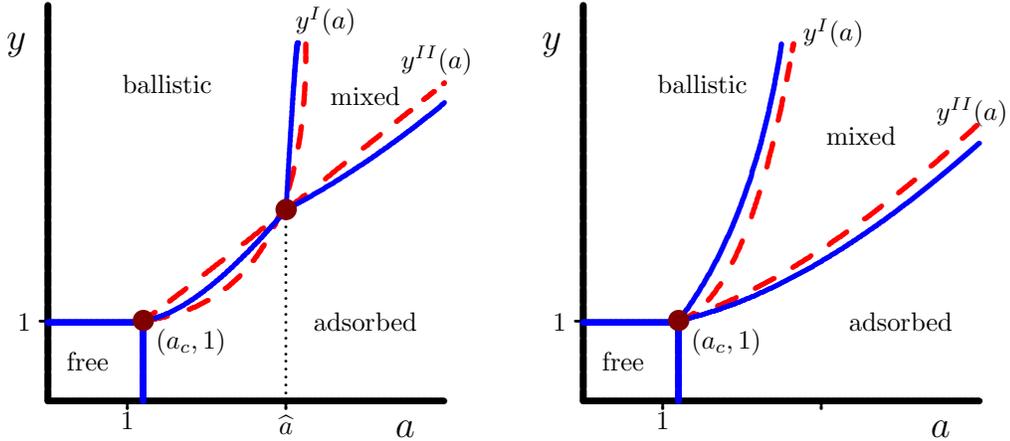

\beginpicture
\setcoordinatesystem units <1.5pt,1.5pt>
\setplotarea x from -40 to 100, y from -10 to 100

\color{black}
\setdots <3pt>
\setplotsymbol ({.})
\plot 60 48 60 0 /
\setsolid
\setplotsymbol ({$\cdot$})
\plot -2 20 0 20 /   \plot 20 -2 20 0 / \plot 60 0 60 -2 /

\setsolid

\setplotsymbol ({\tiny$\bullet$})
\plot 0 100 0 0 100 0 /

\color{black}
\put {\Large$y$} at -8 90
\put {\Large$a$} at 90 -7
\put {$1$} at -6 20
\put {$1$} at 20 -5
\put {$\widehat{a}$} at 60 -6
\put {$y^{I}(a)$} at 70 96
\put {$y^{II}(a)$} at 98 86
\put {$(a_c,1)$} at 36 15

\put {\hbox{free}} at 10 10  
\put {\hbox{ballistic}} at 30 80 
\put {\hbox{adsorbed}} at 80 20 
\put {\hbox{mixed}} at 80 77

\color{blue}
\plot 24 0 24 20 0 20 /

\setplotsymbol ({\LARGE$\cdot$})
\color{red}
\setquadratic
\setdashes <8pt>
\plot 24 20 55 40 65 90  /
\plot 24 20 50 40 100 80 /
\setsolid
\color{blue}
\plot 24 20 40 28 60 48  /
\plot 60 48 80 60 100 75 /
\plot 60 48 62 80 63 90 /
\setlinear
\color{Maroon}
\put {\huge$\bullet$} at 24 20 
\put {\huge$\bullet$} at 60 48

\color{black}
\normalcolor

\setcoordinatesystem units <1.5pt,1.5pt> point at -135 0 
\setplotarea x from -40 to 100, y from -10 to 100

\color{black}
\setplotsymbol ({$\cdot$})
\plot -2 20 0 20 /   \plot 20 -2 20 0 / \plot 60 0 60 -2 /

\setsolid

\setplotsymbol ({\tiny$\bullet$})
\plot 0 100 0 0 100 0 /

\color{black}
\put {\Large$y$} at -8 90
\put {\Large$a$} at 90 -7
\put {$1$} at -6 20
\put {$1$} at 20 -5
\put {$y^{I}(a)$} at 63 93
\put {$y^{II}(a)$} at 98 73
\put {$(a_c,1)$} at 36 15

\put {\hbox{free}} at 10 10  
\put {\hbox{ballistic}} at 30 80 
\put {\hbox{adsorbed}} at 80 20 
\put {\hbox{mixed}} at 70 67

\color{blue}
\plot 24 0 24 20 0 20 /

\setplotsymbol ({\LARGE$\cdot$})
\color{red}
\setquadratic
\setdashes <8pt>
\plot 24 20 60 38 100 70  /
\plot 24 20 40 40 53 90 /
\setsolid
\color{blue}
\plot 24 20 60 35 100 65 /
\plot 24 20 40 50 50 90 /
\setlinear
\color{Maroon}
\put {\huge$\bullet$} at 24 20 

\color{black}
\normalcolor

\endpicture
\caption{If $t<\sfrac{1}{2}$ the basic possible forms of the phase diagram of 
adsorbed self-avoiding walks pulled at an interior vertex are given by these two
panels. If the bounds $y^I(a)$ and $y^{II}(a)$ cross at
$a=\widehat{a}<a_0(t)$ then we do not know that there is a mixed phase for
$a<\widehat{a}$, and the phase diagram may have the form on the left, where there is a 
first order phase boundary between the adsorbed and ballistic phases.  For
$a>\widehat{a}$ there is a mixed phase.  If there is no intersection
between $y^I(a)$ and $y^{II}(a)$ for any value of $a>a_c$, then the phase
diagram will be similar to the diagram on the right.  In this case there is
a mixed phase for all $a>a_c$.  The dashed curves correspond to the bounds
$y^I(a)$ and $y^{II}(a)$, while the phase boundaries are denoted by solid
curves.  If $t\geq \sfrac{1}{2}$ then the curves $y^I(a)$ and $y^{II}(a)$ are reversed
and there is no mixed phase between the adsorbed and ballistic phases.}
\label{fig:phase diagrams}
\end{figure}

There are two basic possible forms that the phase diagram might take, and these are
sketched in Fig.~\ref{fig:phase diagrams}.  In the left hand figure
the curves $y^I(a)$ and $y^{II}(a)$ cross for some $a= \widehat{a} < a_0(t)$.  For values
of $a > \widehat{a}$ there is a mixed phase but for $a< \widehat{a}$ we do not know whether or
not a mixed phase exists.  In the right hand figure
$y^I(a) > y^{II}(a)$ for all $a > a_c$ and a mixed phase exists for all $a>a_c$.

\section{Monte Carlo Results}
\label{sec:montecarlo}

The Monte Carlo simulations of this system are carried out using the flatPERM algorithm \cite{Prellberg2004}.
Self-avoiding walks up to length $n=256$ are grown from a point on the surface defining the half-space of the square and simple cubic lattices.
At each growth step the algorithm records the number of contacts with the surface $v$ and the height above the surface $h$ of the point labelled $\lfloor tn\rfloor$.
FlatPERM produces a flat histogram where every value of $(n,v,h)$ is sampled equally.
For this application, a slight modification is required to sample chains based on the height of an interior vertex.

The normal flatPERM process grows samples in order to produce a histogram that is flat with respect to each microcanonical parameter, however this does not work for the height of an interior vertex.
In this case the chain is not induced to grow away from the surface in the initial stage to achieve a large value of $h$, and then subsequently grow back towards the surface to achieve large $v$.
Thus there is substantial undersampling of configurations that are dominant in the mixed phase.
This issue is exacerbated as $t$ or $n$ increases.
The resolution is to grow the chains normally up to length $\lfloor tn\rfloor$, using a histogram of samples marked by the height of the endpoint vertex $h_\text{end}$ to run the algorithm. 
This ensures that the simulations includes samples that have most of the first $\lfloor tn\rfloor$ vertices extended away from the surface.
When the chain has grown longer than $\lfloor tn\rfloor$, the flattening with respect to the endpoint height is turned off and the histogram is only flattened with respect to $n$ and $v$.
Meanwhile, a second histogram is used to record the samples and weights of each chain with respect to the desired parameters $v$ and $h$ for all chain lengths up to $n$.  
This histogram is used to calculate the correct thermodynamic quantities of the system of interest but is not used in the sampling process.
The benefit of this modified version of flatPERM is to efficiently sample SAWs with respect to the height of the specified interior vertex.

The output of the simulation is the weights $\scw_{nvh}$ that approximate the counts $w_n^{(t)}(v,h)$ used to construct the partition function Eq.~\eqref{eq:PartitionFunction}.
Then we calculate the order parameters $\avm/n$ and $\avh/n$ as weighted sums
\begin{equation}
	\langle Q \rangle_n(a,y) = \frac{\sum_{v,h} Q(n,v,h) \scw_{nvh} a^v y^h }{\sum_{v,h} \scw_{nvh} a^v y^h},
    \label{eq:DoSQuantity}
\end{equation}
where $Q$ is a generic thermodynamic quantity.
Finally, we calculate the Hessian matrix of the finite-size free energy
\begin{equation} 
	H_n =
	\begin{pmatrix}
	 \Sfrac{\partial^2 \fsfree}{\partial a^2} 	& \Sfrac{\partial^2 \fsfree}{\partial a \partial y}	\\
	 \Sfrac{\partial^2 \fsfree}{\partial y \partial a} & \Sfrac{\partial^2 \fsfree}{\partial y^2}
	\end{pmatrix}
	,
	\label{eq:Hessian}
\end{equation}
where derivatives of $\fsfree=\sfrac{1}{n} \log W_n^{(t)}(a,y)$ are calculated using first and second moments of $v$ and $h$ according to Eq.~\eqref{eq:DoSQuantity}.
For each value of $t$ we ran five independent simulations and averaged the results, obtaining a total of $1.3\times 10^{11}$ samples at maximum length $n=256$ on the square lattice and $1.5\times 10^{11}$ samples at maximum length $n=256$ on the simple cubic lattice.

\subsection{Phase diagram}
\label{sec:PhaseDiagram}

\begin{figure}[t!]
\centering
	\includegraphics[width=\textwidth]{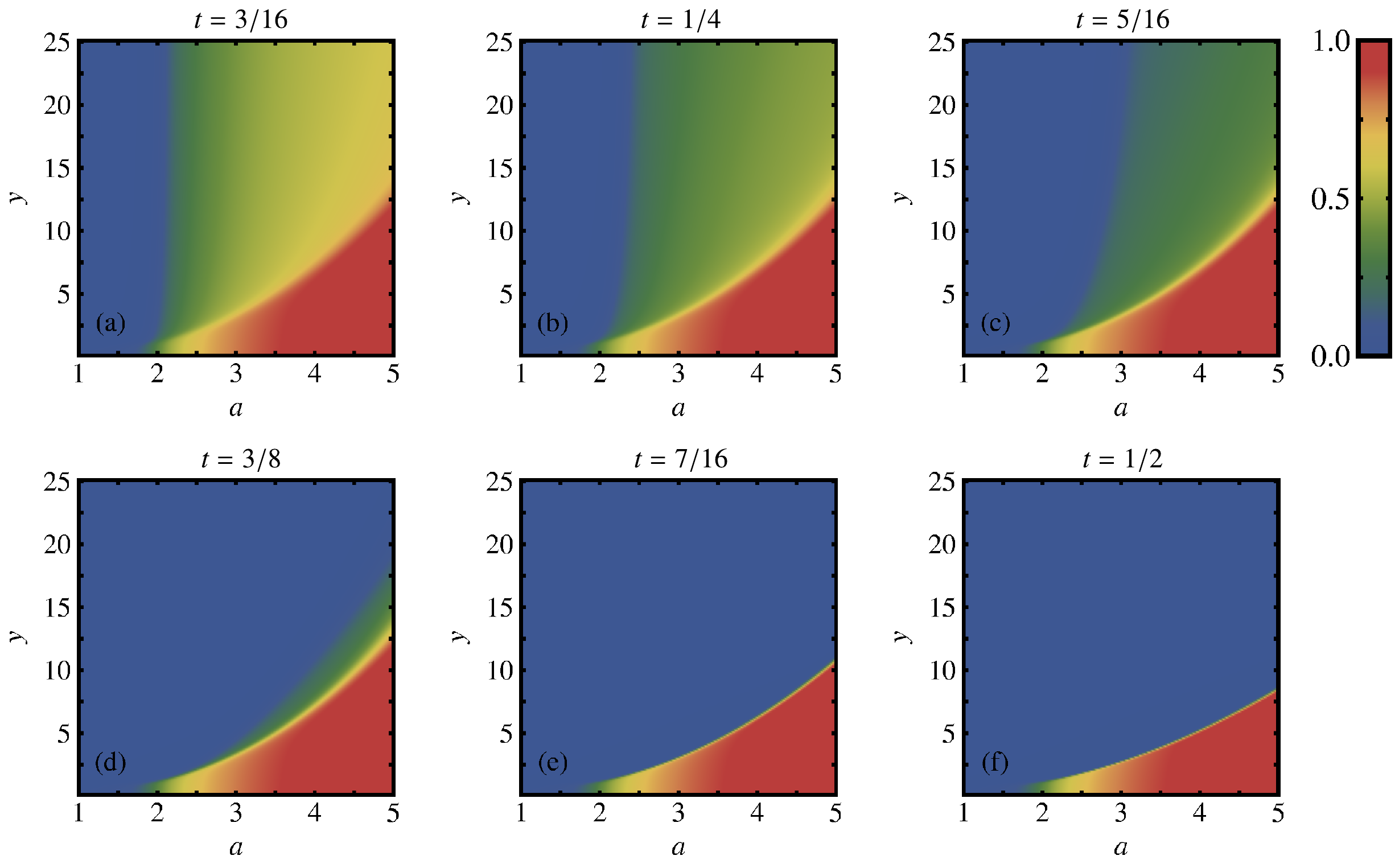}
	
	\caption{Order parameter $\avm/n$ for $n=256$ for a range of $t\le 1/2$ on the square lattice. In each plot the ballistic phase (top left) is distinguished by $\avm/n\approx 0$ (blue), the adsorbed phase (bottom) is distinguished by $\avm/n\approx1$ (red) and the mixed phase (top right), where it occurs, is distinguished by intermediate values of $\avm/n \approx 1-2t$ (green/yellow).
	}
	\label{fig:Order2D_v}
\end{figure}

\begin{figure}[t!]
\centering
	\includegraphics[width=\textwidth]{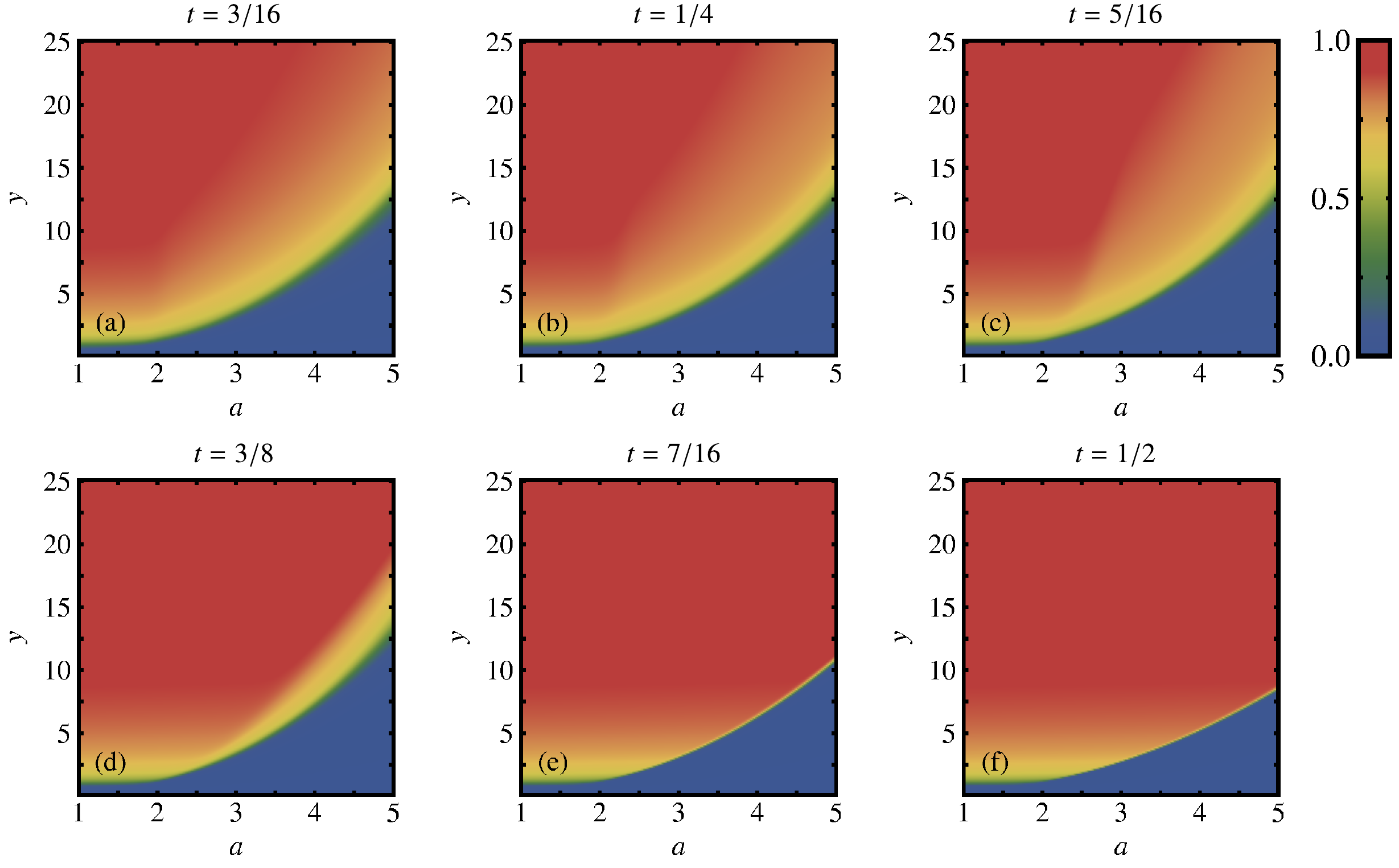}
	
	\caption{Order parameter $\avh/tn$ for $n=256$ for a range of $t\le 1/2$ on the square lattice. In each plot the adsorbed phase (bottom) is distinguished by $\avh/tn \approx 0$ (blue) but there is less distinction between the ballistic and mixed phases (top) where $\avh/tn \approx 1$ (red) for both.
	}
	\label{fig:Order2D_h}
\end{figure}

We focus first on the two-dimensional case, SAWs simulated on the square lattice.
The phase diagram of this system is shown by the order parameters as functions of Boltzmann weights $a$ and $y$.
Figure \ref{fig:Order2D_v} shows the average number of adsorbed vertices $\avm/n$ scaled by chain length $n$ and Fig.~\ref{fig:Order2D_h} shows the average height of the pulled vertex $\avh/tn$ scaled by $tn$, for $t=3/16, \ldots, 1/2$. 
Note that both quantities are on the same colour scale where blue corresponds to 0 and red corresponds to 1.
Collectively, these quantities show the four phases: free, adsorbed, ballistic and mixed.

The free phase is bounded by the adsorption transition point at $a = a_\text{c}$ and the ballistic transition at $y = 1$, and is the region where the surface interaction is absent or repulsive and the force changes from a pull away from the surface into a local push towards the surface. 
This matches the known result for SAWs pulled at the endpoint \cite{Beaton2015}.
Within this phase both the expected number of surface contacts and the average height of the pulled (or pushed in this case) vertex is zero. 

As $a$ increases while keeping $y\leq 1$ the system undergoes a transition to the adsorbed phase at a critical temperature $a_\text{c}>1$.
Beyond the critical point, the average number of surface contacts  quickly approaches its maximum value $\avm/n \approx 1$ (red) while the height of the pulled vertex is suppressed to $\avh/tn \approx 0$ (blue) so almost the entire SAW is adsorbed at the surface.

For $a < a_\text{c}$, as $y$ increases the system enters the ballistic phase at $y=1$, where the thermodynamics depends only on the pulling force.
This phase is characterised by $\avm/n$ tending to zero (blue) while $\avh/tn \approx 1$ (red).
The expected configuration is that the first $tn$ vertices are stretched out away from the surface and then the remainder of the chain assumes a disordered coil configuration relative to the pulled vertex.

For some values of $t$ the mixed phase is visible between the adsorbed and ballistic phases and here $\avm/n<1$ and $\avh/tn<1$ (yellow/green).
This indicates a configuration where the first $tn$ vertices are extended away from the surface, the next $tn$ vertices extend back down to the surface and the remaining $(1-2t)n$ vertices are adsorbed to the surface.
As $t$ increases the mixed phase shrinks as the system tends towards that of a SAW pulled at the midpoint which does not have a mixed phase.

\begin{figure}[t!]
\centering
	\includegraphics[width=\textwidth]{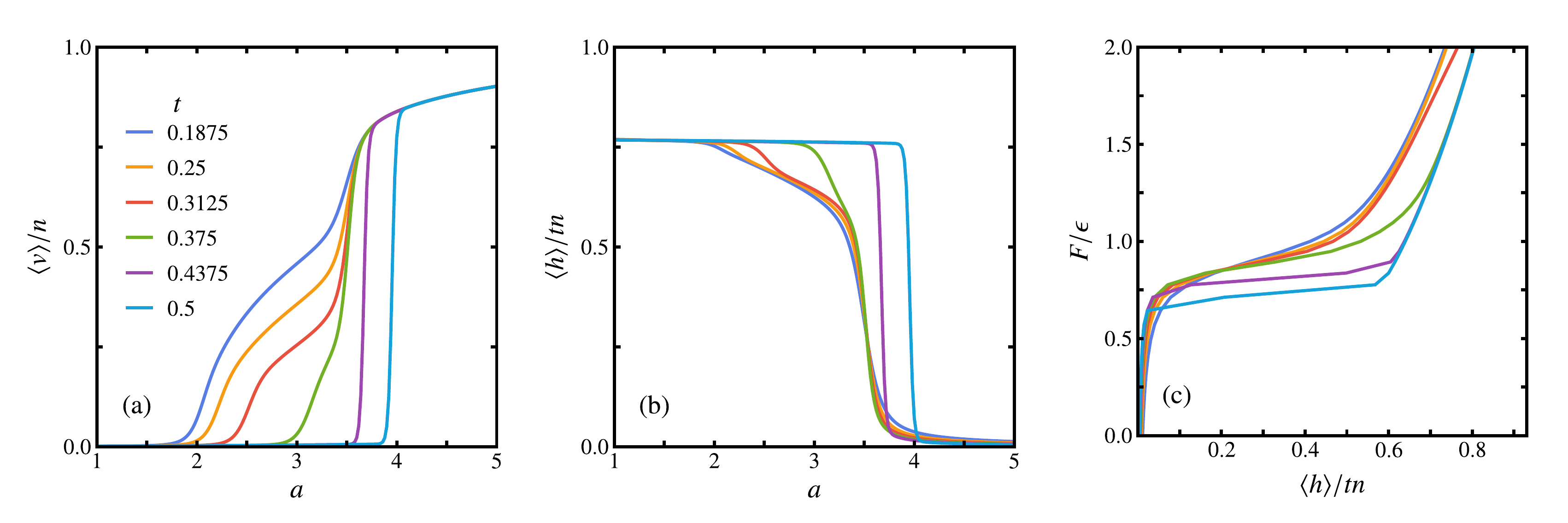}
	
	\caption{Order parameters (a) $\avm/n$ and (b) $\avh/tn$ as functions of $a$ at fixed $y=5.1$ and (c) force extension curves at fixed $a=2.6$.
	Data for all values of $t$ are shown for $n=256$ on the square lattice. 
	}
	\label{fig:OrderLinePlots}
\end{figure}

For a closer look at the scale of the phase transitions we show in Figure \ref{fig:OrderLinePlots} (a) $\langle v \rangle/n$ and (b) $\langle h \rangle /tn$ as functions of $a$ at fixed $y = 5.1$, as well as (c) the force-extension curves at fixed $a=2.6$.  
All values of $t$ are shown, for $n=256$.  
The plots of $\langle v \rangle /n$ for small $t$ clearly show two regions where the number of visits increases rapidly, corresponding to the transitions from the ballistic to the mixed phase and from the mixed to the adsorbed phase.  
These two regions become closer together as $t$ increases and become a single region at $t=1/2$ where there is no mixed phase.  
The values of $\langle h \rangle /tn$ decrease as $a$ increases beyond the ballistic-mixed boundary and then decrease more sharply at the mixed-adsorbed boundary becoming close to zero in the adsorbed phase.  	
The force-extension curves show a single plateau corresponding to the adsorbed-mixed transition.
As $t$ increases towards $1/2$ the plateau becomes more pronounced but the location of the plateau changes only slightly.
There is no plateau corresponding to the mixed-ballistic transition since this is not associated with a major change in the extension at the vertex at which the force is applied.

\subsection{Phase boundaries}
\label{sec:PhaseBoundaries}

\begin{figure}[t!]
\centering
	\includegraphics[width=\textwidth]{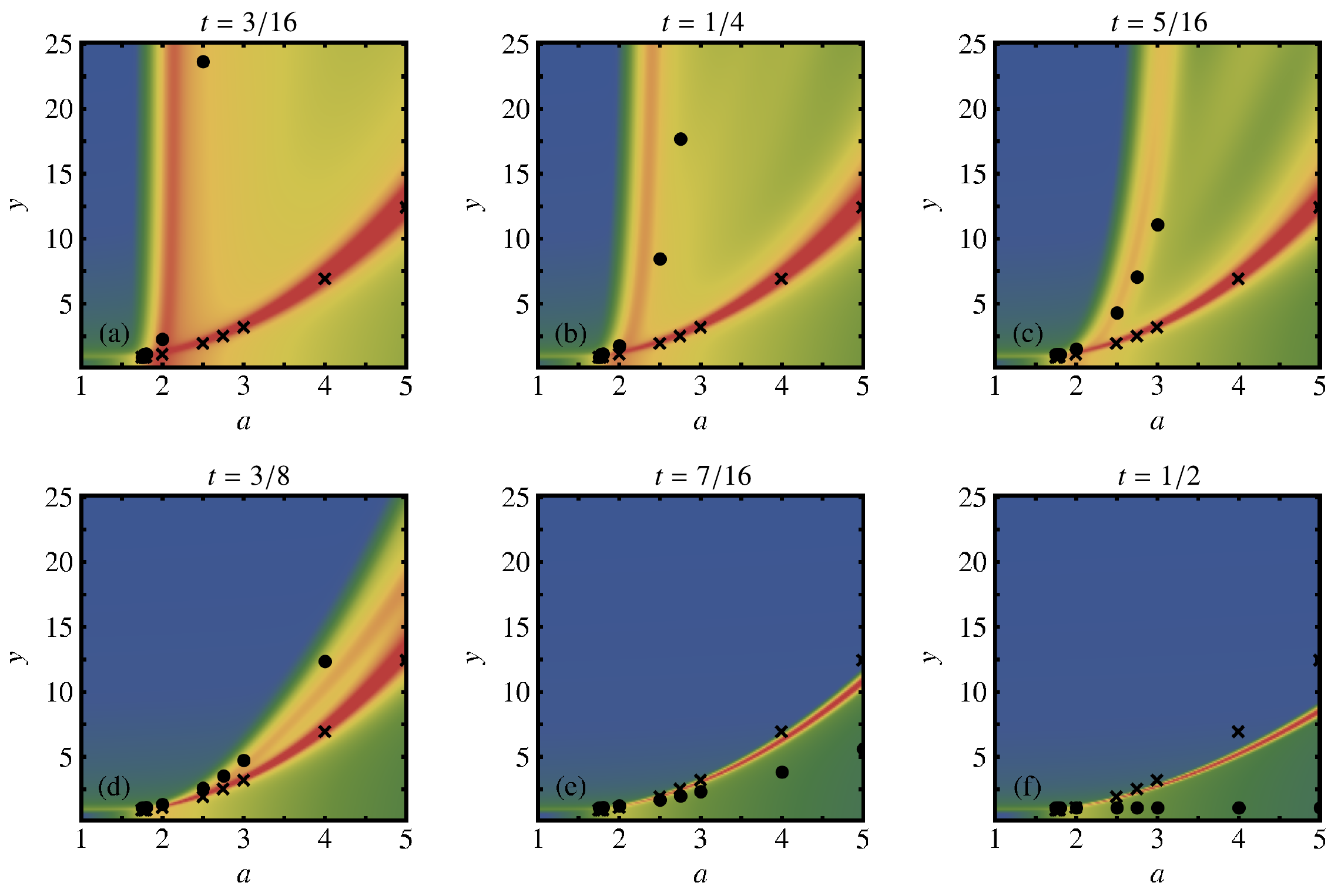}
	
	\caption{Density plot of the logarithm of the largest eigenvalue of the Hessian matrix of the free energy of SAWs on the square lattice for $n=256$ and a range of $t\le 1/2$. 
	Circles and crosses mark points along $y^I(a)$ and $y^{II}(a)$, respectively, calculated from exact enumeration data.
	Red indicates regions of high variance of the order parameters.
	}
	\label{fig:Hessian2D}
\end{figure}

To investigate the phase boundaries more closely we consider the variance of the order parameters, in the form of the Hessian covariance matrix, Eq.~\eqref{eq:Hessian}.
In Fig.~\ref{fig:Hessian2D} we show density plots of the logarithm of the largest eigenvalue of the Hessian matrix of the free energy for $n=256$ and $3/16 \le t\le 1/2$.
In this figure blue corresponds to very small variance and red corresponds to high variance and thus mark the phase boundaries, but are not on a uniform scale.
Overlaid on these plots are the bounding curves $y^I(a)$ and $y^{II}(a)$ defined in Section \ref{sec:rigorous}.
These curves are determined from Eqs.~\eqref{eq:conditionI} and \eqref{eq:conditionII} by using exact enumeration data from Ref.~\cite{Guttmann2014} to calculate $\kappa(a)$ and $\lambda(y)$.
We have previously used this technique to test phase diagrams from Monte Carlo data for a similar problem involving branched polymers \cite{Bradly2019b}.

For the smaller values of $t$ the condition $y^I>y^{II}$ holds and therefore $y^I(a)$ and $y^{II}(a)$ are valid bounds on the ballistic-mixed and adsorbed-mixed phase boundaries.
In Figs.~\ref{fig:Hessian2D}(a-c), corresponding to $t=3/16,1/4,5/16$, the ballistic-mixed boundary lies above $y^I$ and the adsorbed-mixed boundary coincides with $y^{II}$.
At some value of $t$ between $3/8$ and $7/16$ the bounding lines cross over so $y^I<y^{II}$, and thus $y^I$ and $y^{II}$ are no longer bounds on the phase boundaries.
This is reflected in Fig.~\ref{fig:Hessian2D}(e) and (f) where the phase boundaries appear to have merged and lie below $y^{II}$.
Finally, for $t=1/2$ the pulling is at the midpoint so as expected the mixed phase disappears as the phase boundaries merge completely.
It remains an open question as to whether the mixed phase does exist at $t=7/16$ but is vanishingly small for all values of $a$ down to $a_\text{c}$.
Note that $y^I$ is still bounded to the right by the asymptote $a_0$ which for $t=7/16$ is $a_0 \approx 78.7$, so at large enough $a$ the bounds will cross back and $y^I>y^{II}$ again.
However, the enumeration data does not extend to this regime and we do not see the mixed phase re-emerge in the Monte Carlo data.
The value of $t$ where this crossover occurs is not expected to have any physical meaning, and simply reflects how tight the bound is in Eq.~\eqref{eq:lowerbounds}.

The outlier is $t=3/8$, shown in Fig.~\ref{fig:Hessian2D}(d), where $y^I>y^{II}$ but the Monte Carlo data suggests $y^{BM}<y^I$.
We believe this is mainly due to finite-size effects in the Monte Carlo simulations such that the location of the ballistic-mixed boundary $y^{BM}$ for finite $n$ deviates from its value in the thermodynamic limit.
Calculating thermodynamic quantities in the mixed phase (or on its boundaries) is most sensitive to sampling of configurations with both large $h$ and moderately large $v$, which are the hardest to obtain, especially as $t$ increases.
Thus the deviation is exacerbated only for values of $t$ where the phase boundaries are close yet should not be merged, i.e~$t=3/8$.
Note that the adsorbed-mixed boundary still coincides with $y^{II}$ here.
We therefore conclude that the phase diagrams generated from numerical simulation generally agree with the rigorous results.

\begin{figure}[t!]
\centering
	\includegraphics[width=\textwidth]{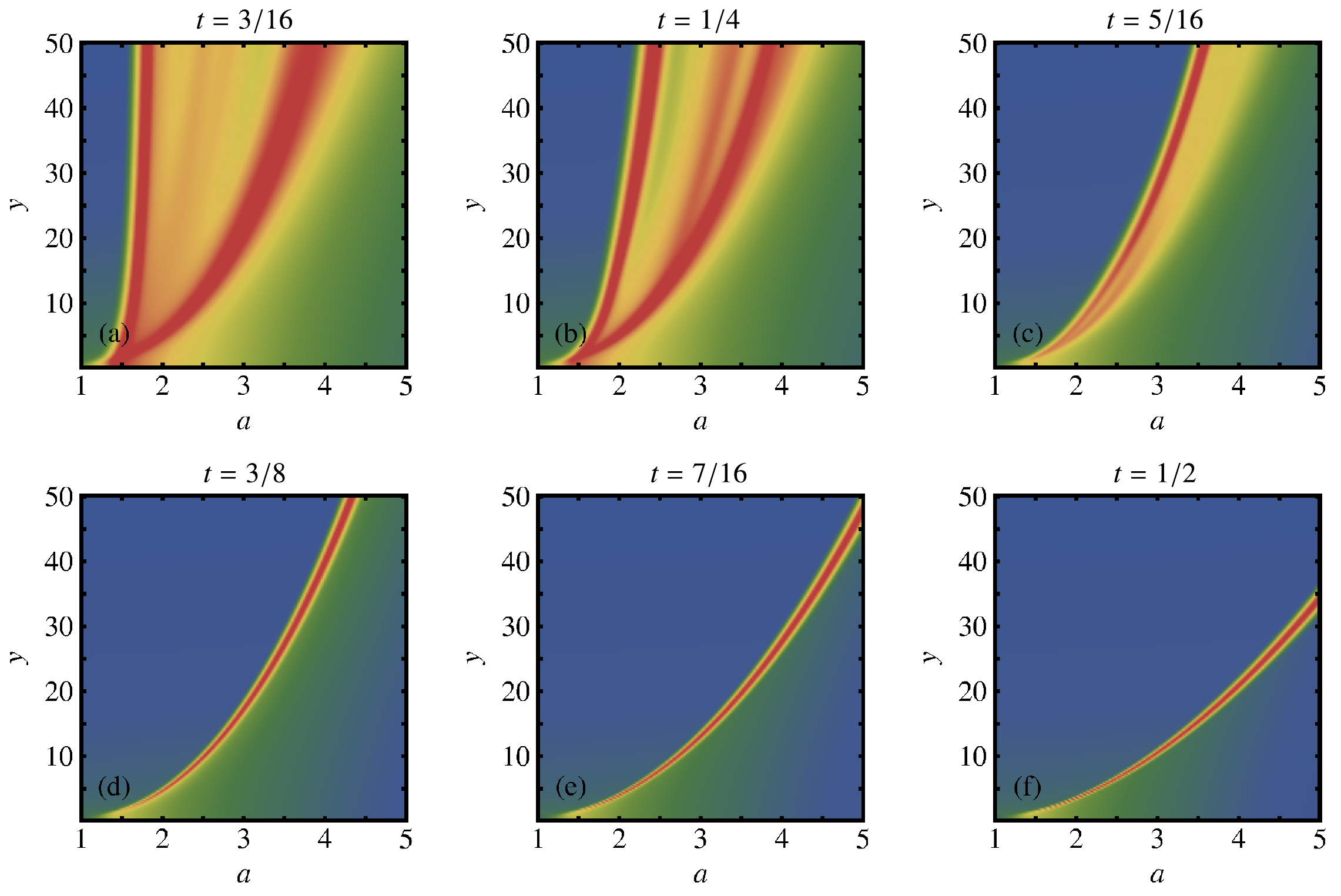}
	
	\caption{Same as Fig.~\ref{fig:Hessian2D} but for SAWs on the simple cubic lattice and a larger range in $y$ on the vertical axis.
	}
	\label{fig:Hessian3D}
\end{figure}

We can also discuss some properties of the system in the limit $t\to 0$, as informed by the trends visible in Figs.~\ref{fig:Order2D_v}, \ref{fig:Order2D_h}, \ref{fig:OrderLinePlots}, and \ref{fig:Hessian2D} for the smaller values of $t$. 
We are already aware that at the point $y=1$ the force is zero and there is only the free and adsorbed phases.
Similarly, in the case $t=0$, for any $y$, the pulling force is applied to the fixed end vertex of the chain and has no effect.
Thus as $t \to 0$ we expect the phase diagram to change to reflect the decreasing $y$-dependence of the free energy.
All four phases will exist for $t>0$ but as $t$ decreases the order parameters in the mixed phase tend toward their values in the adsorbed phase.
This is visible in Figure \ref{fig:OrderLinePlots}(a) and (b) where the latent heat of the adsorbed-mixed transition shrinks near $a=3.5$.
The location of the adsorbed-mixed boundary remains as $t \to 0$ but the ballistic-mixed boundary will become more vertical since it is bounded by $y^I$ and therefore by $a_0$.
By Eq. 13 as $t \to 0$ we obtain $\kappa(a_0(0))=\log\mu_d$, which implies $a_0(0)=a_\text{c}$ since $\kappa(a)$ is strictly increasing for $a>a_\text{c}$.
As $t \to 0$ the mixed phase merges with the adsorbed phase and the ballistic phase merges with the free phase and the free energy is independent of $y$.
This is in accordance with Lemma 1.
The orders of the transitions do not change as $t\to 0$.

Finally we present some results for SAWs pulled at an interior vertex on the simple cubic lattice.
All the results of Section \ref{sec:rigorous} apply generally, namely the existence of all phases and the bounds on their locations.
Each phase in three dimensions is characterised by the same values of the order parameters as the two-dimensional system.
The difference is that the free energies $\kappa(a)$ and $\lambda(y)$ are different functions and we do not have exact enumeration data for SAWs in three dimensions to calculate $y^I$, $y^{II}$ or $a_0$.
Therefore, we show in Fig.~\ref{fig:Hessian3D} density plots of the logarithm of the largest eigenvalue of the Hessian matrix of the free energy for $n=256$ and $3/16 \le t\le 1/2$.
Qualitatively the phase diagrams are the same as the two-dimensional case.
Namely, the mixed phase is apparent at small $t$ with the ballistic-mixed boundary bounded by a vertical asymptote.
As $t$ increases the ballistic-mixed boundary moves towards the adsorbed-mixed boundary and there is some value of $t$ where the two appear to merge or become very close together.
The adsorbed-mixed boundary is independent of $t$ until the ballistic-mixed boundary merges with it.
Without precise knowledge of the free energies in three dimensions we cannot judge how well the boundaries are bounded by the curves $y^I(a)$ and $y^{II}(a)$ but we note that the value of $t$ where the mixed phase disappears is less than in two dimensions.

\section{Discussion}
\label{sec:discussion}

A self-avoiding walk, terminally attached to an impenetrable surface
at which it can adsorb, can be pulled off the surface by applying a force
normal to the surface.  This force can be applied at a particular vertex and,
in this paper, we are concerned with the situation where the force is 
applied at a vertex between the point of attachment and the middle vertex
along the walk.  In this case, even if the walk is completely extended at this vertex,
the remainder of the walk can return to the surface and be partially adsorbed.  

We have examined this situation rigorously and we showed that there are four 
phases, a free phase where the adsorption and the force play little role, an
adsorbed phase, a ballistic phase, and a mixed phase where the free energy
depends on both the force and the strength of the interaction with the surface.
We have derived bounds on the locations of the phase boundaries 
between the ballistic and mixed phases and between the 
mixed and adsorbed phases.  These bounds depend on the vertex at 
which the force is applied.

We have used Monte Carlo methods to map out the details of the phase
diagram as a function of where the force is applied, and we have investigated 
the nature of the phase transitions.  Overall the agreement between the 
Monte Carlo results and the rigorous bounds is excellent.

\section*{Acknowledgement}
C.~B.~thanks York University, Toronto for hosting while part of this work was carried out, 
as well as funding from the University of Melbourne's ECR Global Mobility Award.
Financial support from NSERC of Canada (Discovery Grant RGPIN-2014-04731) and 
from the Australian Research Council via its Discovery Projects scheme (DP160103562) 
is gratefully acknowledged by the authors.


\end{document}